\documentstyle[aps,prl,multicol,epsf,epsfig,times]{revtex}

\date{\today}

\begin{document}
\draft
\title{Three-spin interactions in optical lattices and criticality in cluster Hamiltonians}
\author{Jiannis K. Pachos${}^{1}$ and Martin B. Plenio${}^{2}$ }
\address{
${}^{1}$ Department of Applied Mathematics and Theoretical
Physics, University of Cambridge, Cambridge CB3 0WA, UK,
\\ ${}^{2}$ Quantum Optics and Laser Science Group, Blackett
Laboratory, Imperial College, London SW7 2BW, UK.
 }

\maketitle
\begin{abstract}

We demonstrate that in a triangular configuration of an optical
lattice of two atomic species a variety of novel spin-$1/2$
Hamiltonians can be generated. They include effective three-spin
interactions resulting from the possibility of atoms tunneling
along two different paths. This motivates the study of ground
state properties of various three-spin Hamiltonians in terms of
their two-point and n-point correlations as well as the
localizable entanglement. We present a Hamiltonian with a finite
energy gap above its unique ground state for which the localizable
entanglement length diverges for a wide interval of applied
external fields, while at the same time the classical correlation
length remains finite.

\end{abstract}


\begin{multicols}{2}

\vspace{0.2cm}

The combination of cold atom technology with optical lattices
\cite{Raithel,Mandel} gives rise to a variety of possibilities for
constructing spin Hamiltonians \cite{Kuklov,Duan}. This is
particularly appealing as the high degree of isolation from the
environment that can be achieved in these systems allows for the
study of these Hamiltonians under idealised laboratory conditions.
In parallel, techniques have been developed for minimising
imperfections and impurities \cite{Cirac1,Carl} in the
implementation of the desired structures and for their subsequent
probing and measurement \cite{Roberts}. These achievements permit
the experimental investigation of Hamiltonians that are of
interest in areas such as quantum information or condensed matter
physics with the added advantage of a remarkable freedom in the
choice of external parameters. Presently, attention both in
condensed matter physics and in cold atom research is focusing on
two-spin interactions as these are most readily accessible
experimentally. However, the unique experimental capability
provided by cold atom technology allows us to relax this
restriction. Here we demonstrate that cold atom technology
provides a laboratory to generate and study higher order effects
such as three-spin interactions that give rise to unique
entanglement properties.

The present work serves two purposes. Firstly, it demonstrates
that in a two species Bose-Hubbard model in a triangular
configuration a wide range of Hamilton operators can be generated
that include effective three-spin interactions. They result from
the possibility  of atomic tunneling  through different paths from
one vertex to the other. This can be extended to a one dimensional
spin chain with three-spin interactions. Secondly, we take this
novel experimental capability as a motivation to study unique
ground state properties of Hamiltonians that include three-spin
interactions. In this context one can study possible quantum phase
transitions by considering both the classical correlation
properties as well as the entanglement properties of these
systems. Specifically we consider the so-called cluster
Hamiltonian and its ground state, the cluster state which has
previously been shown to play an important role as a resource in
the context of quantum computation \cite{Briegel R 99}. Subject to
an additional Zeeman term the combined Hamiltonian possesses a
finite energy gap above its unique ground state in a finite
parameter range, hence exhibiting no critical behaviour in the
classical correlations in that regime. We shall show that at the
same time it exhibits a critical behaviour in its entanglement
properties due to its three spin-1/2 interaction term. This is
manifested by a diverging entanglement length of the localizable
entanglement \cite{Verstraete PC 03}. Our example demonstrates
that divergence in entanglement properties are not necessarily
related to the existence of classical critical points, the latter
giving a rather incomplete description of the long-range quantum
correlations against popular belief \cite{Sachdev}. A related
example was arrived at independently in \cite{Verstraete MC 03}.

Consider an ensemble of ultracold bosonic atoms confined in an
optical lattice formed by several standing wave laser beams
\cite{Kuklov,Duan,Jaksch}. Each atom is assumed to have two
relevant internal states, denoted with the index $\sigma=a,b$,
which are trapped by independent standing wave laser beams
differing in polarisation. We are interested in the regime where
the atoms are sufficiently cooled and the periodic potential is
high enough so that the atoms will be confined to the lowest Bloch
band and the low energy evolution can be described by the two
species Bose-Hubbard Hamiltonian \cite{Jaksch}. The tunneling
couplings $J^\sigma$ and the collisional couplings $U_{\sigma
\sigma'}$ can be widely varied by adjusting the amplitude of the
lattice laser fields. For the generation of the multi-particle
interactions discussed here we require large collisional couplings
in order to have a significant effect within the decoherence time
of the system. This can be achieved experimentally by Feshbach
resonances \cite{Inouye,Donley,Kokkelmans}, for which first
theoretical \cite{Mies} and experimental \cite{Donley1} advances
are already promising.

Let us begin by considering the case of only three sites in a
triangular configuration (see Figure \ref{chain}) with tunneling
coupling activated between all three of them. We are interested in
the regime where the tunneling couplings are much smaller than the
collisional ones, $ J^\sigma \ll U_{\sigma \sigma'}$ which
corresponds to the Mott insulating phase and we demand that we
have on average one atom per lattice site. Hence, the basis of
states of site $i$ can be defined by $|\!\!\uparrow\rangle \equiv
|n^a=1, n^b=0\rangle$ and $|\!\!\downarrow \rangle \equiv|n^a=0,
n^b=1\rangle$, where $n^a$ and $n^b$ are the number of atoms in
state $a$ or $b$ respectively. It is possible to expand the
resulting evolution generated by the Bose-Hubbard Hamiltonian in
terms of the small parameters $J^\sigma/U_{\sigma \sigma'}$. In an
interaction picture with respect to the collisional Hamiltonian,
$H^{(0)}=\frac{1}{2} \sum _{i \sigma \sigma'} U_{\sigma \sigma'}
a^{\dagger}_{i\sigma}a^{\dagger}_{i\sigma'}a_{i\sigma'}a_{i\sigma}$,
one obtains the effective evolution from the perturbation
expansion up to the third order with respect to the tunneling
interaction, $V=-\sum_{i\sigma} (J^\sigma_{i} a_{i\sigma}^\dagger
a_{i+1 \sigma} +\text{H.c.})$, given by
\begin{equation}
H=-\sum _\gamma {V_{\alpha \gamma} V_{\gamma \beta} \over
  E_\gamma} +  \sum _{\gamma \delta} {V_{\alpha \gamma} V_{\gamma
    \delta} V_{\delta \beta} \over E_\gamma E_\delta}.
\end{equation}
The indices $\alpha$, $\beta$ refer to states with one atom per
site while $\gamma$, $\delta$ refer to states with two or more
atomic populations per site, $E_\gamma$ are the eigenvalues of the
collisional part, $H^{(0)}$, while we neglected fast rotating
terms effective for long time intervals \cite{Pachos}. Written
explicitly in terms of spin operators we obtain
\begin{eqnarray}
    &&H = \sum_{i=1}^3 \Big[ \vec{B} \cdot \vec{\sigma}_i
    +\lambda^{(1)} \sigma^z_i \sigma^z_{i+1} + \lambda^{(2)}
    (\sigma^x_i \sigma^x_{i+1} +\sigma^y_i \sigma^y_{i+1}) \nonumber\\ &&
    +\lambda^{(3)} \sigma^z_i \sigma^z_{i+1} \sigma^z_{i+2} +
    \lambda^{(4)} (\sigma^x_i \sigma^z_{i+1} \sigma^x_{i+2} +
    \sigma^y_i \sigma^z_{i+1} \sigma^y_{i+2}) \Big]. \label{ham1}
\end{eqnarray}
The couplings $\lambda^{(i)}$ are given as an expansions in
${J^\sigma}/U_{\sigma \sigma'}$ by
\begin{eqnarray*}
\lambda^{(1)} = &&- {{J^a}^2 \over U_{aa}} - {{J^b}^2 \over
U_{bb}} - {9 \over 2} {{J^a}^3 \over U_{aa}^2} - {9 \over
2}{{J^b}^3 \over U_{bb}^2} + {1\over 2} {{J^a}^2+{J^b}^2 \over
U_{ab}} \nonumber\\ && +{1 \over 2} {{J^a}^3+ {J^b}^3 \over
U_{ab}^2} +{ 1\over U_{ab}} \big( {{J^a}^3 \over U_{aa}} + {
{J^b}^3 \over U_{bb}} \big),\nonumber\\
\lambda^{(2)}=&&- {J^a J^b \over U_{ab}}\big(1+{J^a \over U_{aa}}
+ { J^b \over U_{bb}} +{3 \over 2} { J^a  + J^b \over U_{ab}}\big)
\nonumber\\ &&- {J^a J^b \over 2} \big( {J^a \over U_{aa}^2} +{J^b
\over U_{bb}^2} \big), \nonumber\\
\lambda^{(3)}=&&- {3 \over 2} {{J^a}^3 \over U_{aa}^2}+ {3 \over
2} {{J^b}^3 \over U_{bb}^2} + {1 \over U_{ab}} \big( {{J^a}^3
\over U_{aa}} - {{J^b}^3 \over U_{bb}} \big),\nonumber\\
\lambda^{(4)}=&& -{{J^a} J^b \over U_{ab}} \big({{J^a} \over
U_{aa}} - {J^b \over U_{bb}} \big)- {J^a J^b \over 2} \big({J^a \over
  U_{aa}^2} -{J^b \over U_{bb}^2} \big).
\nonumber
\end{eqnarray*}
The local field $\vec{B}$ can be arbitrarily tuned by applying
appropriately detuned laser fields while we need to compensate for
single particle phase rotations of the form $B_z \sum_i\sigma^z_i$
with
\begin{displaymath}
B_z = -{{J^a}^2\over U_{aa}}(2+\frac{ 9J^a}{2U_{aa}}+\frac{J^a}{U_{ab}})
    + {{J^b}^2\over U_{bb}}(2+\frac{
    9J^b}{2U_{bb}}+\frac{J^b}{U_{ab}}).
\end{displaymath}
One can isolate different parts from Eq. (\ref{ham1}), each one
including a three-spin interaction term, by varying the tunneling
and/or the collisional couplings appropriately so that particular
$\lambda^{(i)}$ terms such as the two spin interactions vanish,
while others can be varied freely.

By employing additional Raman transitions in such a way as to
couple the states $a$ and $b$ during
tunneling it is possible to obtain variations of the above
Hamiltonian \cite{Duan}. Indeed, Raman transitions can activate
tunneling of the states $|+ \rangle\equiv {1 \over \sqrt{2}}
(|\!\!\uparrow\rangle+|\!\!\downarrow\rangle)$, while
the tunneling of the states $|- \rangle\equiv {1 \over \sqrt{2}}
(|\!\!\uparrow\rangle-|\!\!\downarrow\rangle)$ is obstructed.
Hence, it is possible to generate different
coefficients in front of the two spin interaction terms
$\sigma^x_i \sigma^x_{i+1}$ or $\sigma^y_i \sigma^y_{i+1}$ as they
are the diagonal matrices in their corresponding basis $|+\rangle$
or $|-\rangle$. Considering its effect on the three-spin
interaction it is possible to generate additional terms of the
form $\sigma^x_i \sigma^x_{i+1} \sigma^x_{i+2}$ or $\sigma^y_i
\sigma^y_{i+1} \sigma^y_{i+2}$ with couplings similar to
$\lambda^{(3)}$. Note that the effective spin interactions produced by
Raman transitions {\em do not} preserve the number of particles in
each of the species.

In particular, we are interested in obtaining a whole chain of
triangles in a zig-zag one dimensional pattern as in Fig.
\ref{chain}. Indeed, with this configuration we can extend from
a single triangle to a whole triangular ladder. Nevertheless, a careful
consideration of the two spin interactions shows that terms of the
form $\sigma^z_i \sigma^z_{i+2}$ also appear, due to the triangular
configuration (see Fig. \ref{chain}). Hamiltonians involving nearest
and next-to-nearest neighbour interactions are of interest in their own
right (see e.g. Chapter 14 of \cite{Sachdev} and \cite{Sachdev1}), but
we will not address these systems here. It is possible to introduce a
longitudinal optical lattice with half of the initial wave length, and
an appropriate amplitude such that it cancels exactly those
interactions generating finally chains with only neighbouring
couplings.
\begin{minipage}{3.38truein}
\begin{figure}[ht]
\centerline{
    \epsfig{file=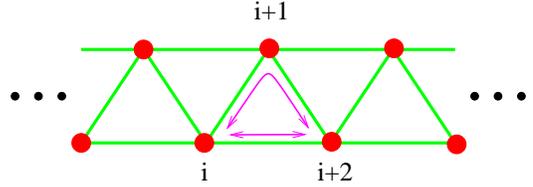,width=2.7truein}
    \put(-123,-10){i}
    \put(-79,-10){i+2}
    \put(-103,51){i+1} }
    \vspace{0.5cm}
    \caption[chain]{\label{chain} The one dimensional chain constructed
    out of equilateral triangles. Three-spin interaction terms appear, e.g. between
    sites $i$, $i+1$ and $i+2$ as, for example, tunneling between $i$ and $i+2$
    can happen through two different paths, directly and through site
    $i+1$. The latter resulting into an interaction between
    $i$ and $i+2$ that is controlled by the state of site $i+1$.}
\end{figure}
\vspace*{-0.2cm}
\end{minipage}

In a similar fashion it is possible to avoid generation of the
term $\sigma^x_i \sigma^x_{i+2} + \sigma^y_i \sigma^y_{i+2}$ by
deactivating the longitudinal tunnelling coupling in one of the
modes, e.g. the $a$ mode which deactivates the corresponding
exchange interaction. We are particularly interested in three-spin
interactions and would like to isolate the chain term $\sum_i
(\sigma^x_i \sigma^z_{i+1} \sigma^x_{i+2}+\sigma^y_i
\sigma^z_{i+1} \sigma^y_{i+2})$ from the $\lambda^{(4)}$ term (see
Hamiltonian (\ref{ham1})) that includes in addition all the
possible triangular permutations. To achieve that we could
deactivate now the non-longitudinal tunnelling for one of the two
modes, e.g. the $a$-mode. With the above procedures we can finally
obtain a chain Hamiltonian as in (\ref{ham1}) where the summation
runs up to the total number $N$ of the sites. A variety of
different Hamiltonians could be generated by different
combinations of the above techniques.

In the past Hamiltonians describing three-spin interactions have
been of limited interest \cite{threespinpapers} as they were
difficult to implement and control experimentally. The above
results demonstrate that Hamiltonians with three-spin interactions
can be implemented and controlled across a wide parameter range.
One may suspect that ground states of three-spin interaction
Hamiltonians exhibit unique properties as compared to ground
states generated merely by two-spin interaction. This motivates
the study of the properties of the ground state of a particular
three-spin Hamiltonian for different parametric regimes. Possible
phase transitions induced by varying these parameters are explored
employing two possible signatures of critical behaviour that are
quite different in nature. In particular, new critical phenomena
in three-spin Hamiltonians that cannot be detected on the level of
classical correlations will be demonstrated.

(i) A traditional approach to criticality of the ground state
studies two-point correlation functions between spins $1$ and $L$,
given by $ C_{1L}^{\alpha \beta} \equiv
    \langle \sigma_1 ^{\alpha} \sigma_L^{\beta} \rangle
    -\langle \sigma_1 ^{\alpha}\rangle
    \langle  \sigma_L^{\beta}\rangle
$, for varying $L$, where $\alpha,\beta=x,y,z$. These two-point
correlations may exhibit two types of generic behaviours, namely
(a) exponential decay in $L$, i.e. the correlation length $\xi$,
defined as
\begin{equation}
    \xi^{-1} \equiv \lim_{L\rightarrow\infty} \frac{1}{L}\log\,
    C_{1L}^{\alpha \beta},
    \label{corlength1}
\end{equation}
is finite or, (b), power-law decay in $L$, i.e. $C_{1L}^{\alpha
\beta}\sim L^{-q}$ for some $q$, which implies an infinite
correlation length $\xi$ indicating a critical point in the system
\cite{Sachdev}.

(ii) While the two-point correlation functions ${\cal
C}_{1L}^{\alpha \beta}$ are a possible indicator for critical
behaviour, they provide an incomplete view of the quantum
correlations between spins $1$ and $L$. Indeed they ignore
correlations through all the other spins by tracing them out.
Already the GHZ state $|GHZ\rangle = (|000\rangle +
|111\rangle)/\sqrt{2}$ shows that this looses important
information. Tracing out particle $2$ leaves particles $1$ and $3$
in an unentangled state. However, measuring the second particle in
the $\sigma_x$-eigenbasis leaves particles $1$ and $3$ in a
maximally entangled state. Therefore one may define the
localizable entanglement $E^{(loc)}_{1L}$ between spins $1$ and
$L$ as the largest average entanglement that can be obtained by
performing optimised local measurements on all the other spins
\cite{Verstraete PC 03}. In analogy to Eq. (\ref{corlength1}) one
can define the entanglement length
\begin{equation}
    \xi_E^{-1} \equiv \lim_{L\rightarrow\infty} \frac{1}{L} \log\,
    E^{(loc)}_{1L}.
    \label{corlength2}
\end{equation}

It is an interesting question whether criticality according to one
of these indicators implies criticality according to the other.
The localizable entanglement length is always larger than or equal
to the two-point correlation length and indeed, it has been shown
that there are cases where criticality behaviour can be revealed
only by the diverging localizable entanglement length while the
classical correlation length remains finite \cite{Verstraete MC
03}. Such behaviour is also expected to appear when we consider
particular three-spin interaction Hamiltonians. To see this
consider the Hamiltonian
\begin{equation}
    H = \sum_{i} \big( -\sigma^x_{i-1}\sigma^z_i\sigma^x_{i+1} +
    B\sigma^z_i \big),
    \label{xzxmodel}
\end{equation}
where we assume periodic boundary conditions. The fact that
$\sigma^x_{i-1}\sigma^z_i\sigma^x_{i+1}$ commute for different $i$
and employing raising operator $L^{\dagger}_k=\sigma^x_k -i
\sigma^x_{k-1} \sigma^y_{k}\sigma^x_{k+1}$ allows to determine the
entire spectrum of $H$ for $B=0$. The unique ground state of $H$
for $B=0$ is the well-known cluster state \cite{Briegel R
99,Verstraete}, which has previously been studied as a resource in
the context of quantum computation. It possesses a finite energy
gap of $\Delta E=2$ above its ground state \cite{comment}. For
finite $B$ the energy eigenvalues of the system can still be found
using the Jordan-Wigner transformation and a lengthy but
straightforward calculation shows that the energy gap persists for
$|B|\neq 1$. The exact solution also shows that the system has
critical points for $|B|=1$ at which the two-point correlation
length and the entanglement length diverges. For any other value
of $B$ and in particular for $B=0$ the system does not exhibit a
diverging two-point correlation length as is expected from the
finite energy gap above the ground state. Indeed, correlation
functions such as
\begin{eqnarray}
\label{zzcorrelations}
    C^{zz}_{1L} &=& \big(\frac{1}{4\pi}\int_{-2\pi}^{2\pi}\!
    \frac{\sin r}{\sqrt{B^2 + 1 + 2B \cos r}}\sin
    \frac{(L-1)r}{2} dr \big)^2  \nonumber\\[0.2cm]
     -&& \!\!\!\!\!\!\!  \big(\frac{1}{4\pi}\int_{-2\pi}^{2\pi}\!
    \frac{B+\cos r}{\sqrt{B^2 + 1 + 2B \cos r}}\cos
    \frac{(L-1)r}{2} dr \big)^2
\end{eqnarray}
can be computed and the corresponding correlation length can be
explicitly determined analytically using standard techniques (see
e.g. Fig. (\ref{entanglementlength})) \cite{Barouch M 71}. The
two-point correlation functions such as Eq.
(\ref{zzcorrelations}) exhibit a power-law decay at the critical
points $|B|=1$ while they decay exponentially for all other values
of $B$ in contrast to the anisotropic $XY$-model whose
$C^{xx}_{1L}$ correlation function tends to a finite constant in
the limit of $L\rightarrow\infty$ for $|B|<1$ \cite{Barouch M 71}.
This discrepancy is due to the finite energy gap the model in Eq.
(\ref{xzxmodel}) exhibits above a non-degenerate ground state in
the interval $|B|<1$.

When we study three-spin interactions it is natural to consider
the behaviour of higher-order correlations. For the ground state
with magnetic field $B=0$ all three-point correlation except,
obviously,
$\langle\sigma^x_{i-1}\sigma^z_{i}\sigma^x_{i+1}\rangle$ vanish.
Indeed, if we consider $n>4$ neighbouring sites and chose for each
of these randomly one of the operators
$\sigma_x,\sigma_y,\sigma_z$ or ${\bf 1}$ then the probability
that the resulting correlation will be non-vanishing is given by
$p=2^{-(2+n)}$. For $|B|>0$ however far more correlations are
non-vanishing and the rate of non-vanishing correlations scales
approximately as $0.858^n$. This marked difference which
distinguishes $B=0$ is due to the higher symmetry that the
Hamiltonian exhibits at that point.

In the following we shall consider the localizable entanglement
and the corresponding length as described in (ii). Compared to the
two-point correlations, the computation of the localizable
entanglement is considerably more involved due to the optimization
process. Nevertheless, it is easy to show that the entanglement
length diverges for $B=0$. In that case the ground state of the
Hamiltonian (\ref{xzxmodel}) is a cluster state with the property
that any two spins can be made deterministically maximally
entangled by measuring the $\sigma_z$ operator on each spin in
between the target spins, while measuring the $\sigma_x$ operator
on the remaining spins. Indeed, this property underlies its
importance for quantum computation as it allows to propagate a
quantum computation through the lattice via local measurements
\cite{Briegel R 99}.

For finite values of $B$ it is difficult to obtain the exact value
of the localizable entanglement. Nevertheless, to establish a
diverging entanglement length it is sufficient to provide lower
bounds that can be obtained by prescribing specific measurement
schemes. Indeed, for the ground state of (\ref{xzxmodel}) in the
interval $|B|<1$ consider two spins $1$ and $L=2k+1$ where $k\in
{\bf N}$. Measure the $\sigma_x$ operator on spin $2$ and on all
remaining spins, other than $1$ and $L$, the $\sigma_z$ operator.
By knowing the analytic form of the ground state one can obtain
the average entanglement over all possible measurement outcomes in
terms of the concurrence, that tends to $ E_{\infty} =
\left(1-|B|^2\right)^{1/4}$ for $k\rightarrow\infty$. This
demonstrates that the localizable entanglement length is infinite
in the full interval $|B|<1$. This surprising critical behaviour
for the whole interval $|B|<1$ is {\em not} revealed by the
two-point or n-point correlation function which exhibit finite
correlation lengths. For $|B|>1$ however, numerical results,
employing a simulated annealing technique to find the optimal
measurement for a chain of 16 spins, show that the localizable
entanglement exhibits a finite length scale.
\begin{minipage}{3.38truein}
\begin{figure}[ht]
\centerline{
    \epsfig{file=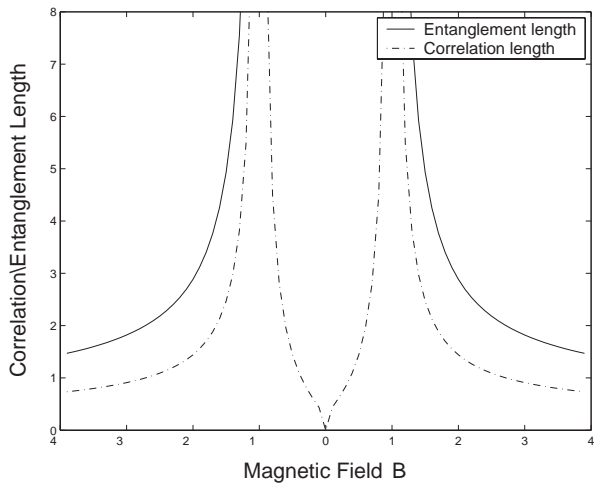,width=3.1truein}
   \vspace{0.5cm}}
    \caption[chain]{\label{entanglementlength} Both, the two-point correlation
    length for $C^{zz}_{1L}$ (solid line) and the localizable entanglement
    length (dashed line) are shown for various  magnetic field for
    chain of length 16. Note that the localizable entanglement
    length diverges in the whole interval $|B|<1$ while the two-point
    correlation length is finite in this interval. }
\end{figure}
\vspace*{-0.4cm}
\end{minipage}

In Fig. (\ref{entanglementlength}) both the two-point correlation
length and localizable entanglement length are drawn versus the
magnetic field. In the interval $|B|<1$ the entanglement length
diverges while the correlation length remains finite. For finite
temperatures the localizable entanglement becomes finite
everywhere but, for temperatures that are much smaller than the
gap above the ground state, it remains considerably larger than
the classical correlation length. This demonstrates the resilience
of this phenomenon against thermal perturbations.

To summarize, we have demonstrated that various Hamiltonians
describing three-spin interactions can be created in triangular
optical lattices in a two-species Bose-Hubbard model. They can be
realized in the laboratory with the near future cold atom
technology. In fact, a study of the required experimental values
reveals that with a tunneling coupling $J/\hbar\sim$10 kHz
\cite{Mandel} an experimentally achievable collisional coupling of
$U/ \hbar \sim$100 kHz is required. With these values a full
numerical study demonstrates that the perturbative truncation is
valid within a $4\%$ error and a significant effect of the three
spin interactions is obtained within the decoherence time of the
system taken here to be $10$ms. Previously, the systematic
experimental creation of three-spin interaction Hamiltonians has
been extremely difficult. The new capability for the systematic
creation of three-spin Hamiltonians and their possible isolation
from other interactions motivates the study of the properties of
their ground states and here in particular of their phase
transitions. Motivated by this we presented a particular
three-spin cluster Hamiltonian that exhibits a novel kind of
critical behavior that is not revealed by two-point correlation
functions. In addition, interactions such as $\sigma^z_1
\sigma^z_2 \sigma^z_3$ presented here have proved to be of
interest for quantum computation. They can implement multi-qubit
gates, like the Toffoli gate, in essentially one step \cite{Pachos
K 03} reducing dramatically the experimental resources.

{\em Acknowledgements.} We thank Derek Lee for inspiring
conversations. This work was supported by a Royal Society
University Research Fellowship, a Royal Society Leverhulme Trust
Senior Research Fellowship, the EU Thematic Network QUPRODIS and
the QIP-IRC of EPSRC.


\end{multicols}
\end{document}